\documentclass{article}
\usepackage{spconf,amsmath,graphicx,hyperref}
\usepackage{enumitem}
\usepackage{spconf,amsmath,graphicx,hyperref}
\usepackage{subcaption} 
\usepackage{fancyhdr}
\usepackage{pifont} 
\usepackage{multirow,makecell}
\usepackage{amsmath,bm,graphicx}
\usepackage{bm} 
\usepackage{subcaption}
\usepackage{amsfonts}
\usepackage{algorithmic,algorithm}
\usepackage{enumitem}
\usepackage{adjustbox} 
\usepackage{booktabs}

\title{CodeSep: Low-Bitrate Codec-Driven Speech Separation with Base-Token Disentanglement and Auxiliary-Token Serial Prediction}
\name{Hui-Peng Du, Yang Ai$^*$\thanks{$^*$Corresponding author.}, Xiao-Hang Jiang, Rui-Chen Zheng, Zhen-Hua Ling}
\address{National Engineering Research Center of Speech and Language Information Processing, \\University of Science and Technology of China, Hefei, P. R. China\\
{\small \tt \ \{redmist,jiang\_xiaohang,zhengruichen\}@mail.ustc.edu.cn, \{yangai,zhling\}@ustc.edu.cn}
}
\begin{document}
\ninept
\maketitle
\begin{abstract}
This paper targets a new scenario that integrates speech separation with speech compression, aiming to disentangle multiple speakers while producing discrete representations for efficient transmission or storage, with applications in online meetings and dialogue archiving. 
To address this scenario, we propose CodeSep, a codec-driven model that jointly performs speech separation and low-bitrate compression. 
CodeSep comprises a residual vector quantizer (RVQ)-based plain neural speech codec, a base-token disentanglement (BTD) module, and parallel auxiliary-token serial prediction (ATSP) modules. 
The BTD module disentangles mixed-speech mel-spectrograms into base tokens for each speaker, which are then refined by ATSP modules to serially predict auxiliary tokens, and finally, all tokens are decoded to reconstruct separated waveforms through the codec decoder. 
During training, the codec’s RVQ provides supervision with permutation-invariant and teacher-forcing-based cross-entropy losses. 
As only base tokens are transmitted or stored, CodeSep achieves low-bitrate compression. 
Experimental results show that CodeSep attains satisfactory separation performance at only 1 kbps compared with baseline methods.

\end{abstract}
\section{Introduction}
\label{sec:intro}

Speech separation, also known as the cocktail party problem \cite{cherry1953some}, aims to isolate individual speakers from a mixed signal containing multiple voices. 
Recent advances in neural networks have led to remarkable progress in this task. 
Architectures based on convolutional neural networks (CNNs) \cite{luo2019conv,tzinis2020sudo,li2022efficient,hu2021speech}, recurrent neural networks (RNNs) \cite{li2022use,li2023design,luo2020dual}, and Transformers \cite{chen2020dual,subakan2021attention,yang2022tfpsnet} have all been applied to speech separation, each achieving state-of-the-art (SOTA) performance upon introduction. 
More recently, state space model (SSM)-based approaches \cite{chen2023neural,li2024spmamba} have emerged, enabling joint long- and short-term modeling of speech through selective spatial processing. 
For evaluation, most methods adopt the scale-invariant signal-to-distortion ratio (SI-SDR) \cite{le2019sdr} as both the training objective and performance metric, ensuring effective and high-quality separation.



However, existing speech separation methods focus solely on separating speakers from a mixture, without considering scenarios where both separation and compression are required. 
In practice, it is often necessary to generate compact discrete representations alongside separation for efficient transmission or storage. 
For instance, in \textit{online meetings}, mixed speech must be separated in real time and represented as discrete tokens to support transcription or meeting summarization. 
In \textit{dialogue archiving}, compressed discrete representations of separated speech help reduce storage cost while preserving accessibility for later retrieval and analysis. 
These scenarios highlight the necessity of jointly addressing speech separation and compression, which is rarely considered in current approaches. 
Recent works such as Codecformer \cite{yip2024towards} and Codecformer-EL \cite{yip2024speech} combine speech separation models with speech codecs \cite{kumar2023high}, providing a new perspective on separation. 
However, they only exploit the codec architecture without producing discrete representations, thus leaving the compression-oriented scenario unaddressed.

\label{sec:pagestyle}
\begin{figure*}[h!]
    \centering
    \includegraphics[width=1\linewidth]{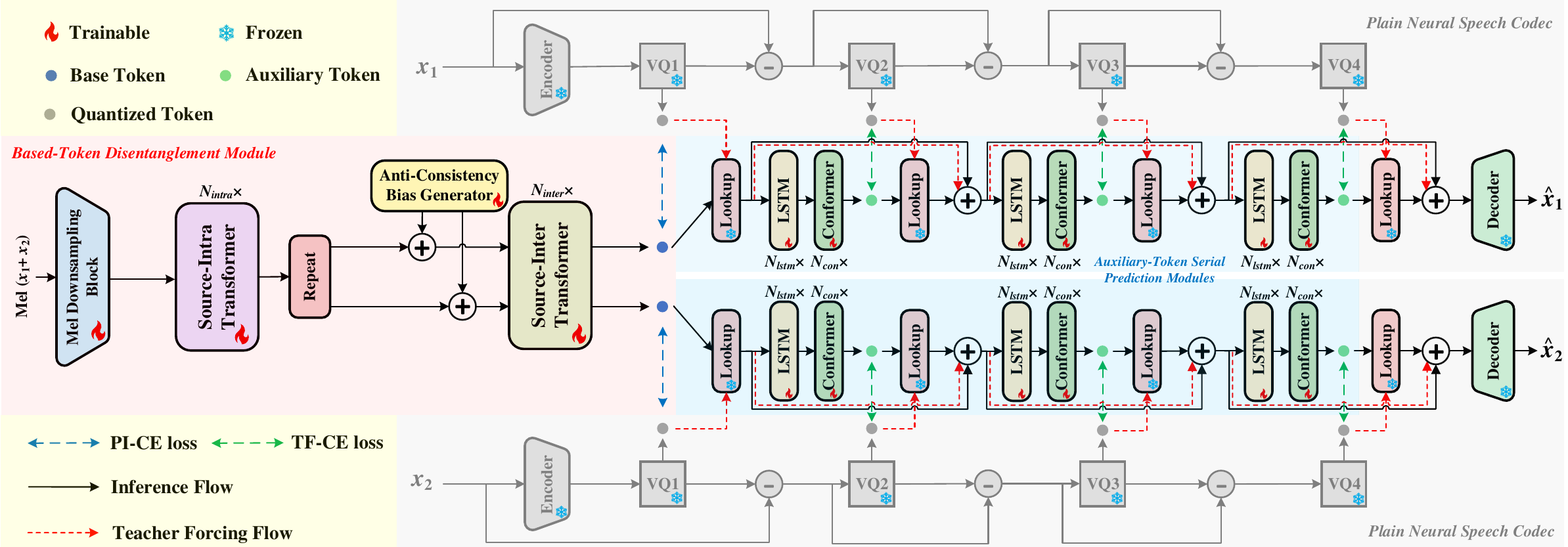}
    \caption{An overview of CodeSep with $N=4$ as an example. The gray parts indicate modules absent during inference.}
    \label{fig:2}
\end{figure*}

\addtolength{\dbltextfloatsep}{-0.2cm}

To address this scenario, we propose CodeSep, which simultaneously achieves speech separation and low-bitrate compression. 
CodeSep is a discrete-domain codec-driven speech separation method, which disentangles speaker information at the level of discrete tokens. 
It relies on a residual vector quantizer (RVQ)-based \cite{zeghidour2021soundstream} plain neural speech codec to provide token targets and decoding. 
Specifically, the base-token disentanglement (BTD) module in CodeSep first takes the mel-spectrogram of the mixed speech as input and generates disentangled base tokens corresponding to each speaker. 
Subsequently, the parallel auxiliary-token serial prediction (ATSP) modules take each disentangled base token and serially predict its auxiliary tokens, after which the codec decoder reconstructs the corresponding speech waveforms. 
In CodeSep, only base tokens are transmitted or stored, allowing the receiver or decompression side to reconstruct the speech and thus achieving low-bitrate compression. 
Both objective and subjective experiments confirm that CodeSep, at only 1 kbps, significantly outperforms baseline methods in terms of separated speech quality and speaker similarity.


\vspace{-3mm}
\section{Task Definition and Solutions}
\vspace{-3mm}
\label{sec: task def}
This work targets a new scenario that integrates speech separation with speech compression, where mixed speech must not only be separated into individual speakers but also discretized for efficient transmission or storage. 
Specifically, the task can be formulated as follows. 
Taking the mixture of dual speech signals as an example, given the mixed speech $ \bm{y} = \bm{x}_1 + \bm{x}_2 $, it must be discretized into tokens for transmission or storage. 
At the receiver or decompression side, the separated speeches $ \hat{\bm{x}}_1 $ and $ \hat{\bm{x}}_2 $ are then reconstructed from these tokens.

For this task, two straightforward solutions can be considered:
\begin{itemize}[leftmargin=*]
\item {\textbf{First-Compression-Then-Separation (FCTS) Solution:}} In this solution, the mixed speech $ \bm{y} $ is first compressed by a speech codec into a discrete representation $ d^{(\bm{y})} $, decoded back into $ \bm{y}' $, and then passed to a speech separation model to produce $ \hat{\bm{x}}_1 $ and $ \hat{\bm{x}}_2 $. 
However, this solution demands a relatively high bitrate, as compressing and reconstructing mixed speech is inherently more challenging than single speech, and separation models are particularly sensitive to input quality.

\item {\textbf{First-Separation-Then-Compression (FSTC) Solution:}} In this solution, the mixed speech $ \bm{y} $ is first separated into $ \bm{x}_1' $ and $ \bm{x}_2' $ by a speech separation model, which are then repectively compressed by a speech codec into discrete tokens $ d^{(\bm{x}_1')} $ and $ d^{(\bm{x}_2')} $, and finally decoded to reconstruct $ \hat{\bm{x}}_1 $ and $ \hat{\bm{x}}_2 $. 
This solution also requires a high bitrate, since both separated speeches must be compressed. 
The bitrate grows proportionally with the number of speakers in the mixture.

\end{itemize}
In contrast, the proposed CodeSep offers a joint-separation-and-compression (JSAC) solution that features both high-quality separation and low-bitrate compression.
The details are presented below.

\vspace{-1mm}
\section{Proposed Method}
\vspace{-1mm}


The architecture of CodeSep is illustrated in Fig. \ref{fig:2}, which consists of an RVQ-based plain neural speech codec, a BTD module, and parallel ATSP modules. 
The codec provides training objectives for the other modules and performs decoding, while the other two modules separate speaker-specific base tokens from the mixture and serially predict the remaining auxiliary tokens. 
All modules of CodeSep are trained independently and integrated during inference.

\vspace{-4mm}
\subsection{Plain Neural Speech Codec}
\vspace{-3mm}
CodeSep is codec-driven, and we adopt our previously proposed MDCTCodec \cite{jiang2024mdctcodec} as the plain neural speech codec. 
The MDCTCodec operates on the modified discrete cosine transform (MDCT) spectrum for encoding, quantization, and decoding, achieving both high-quality and efficient speech reconstruction. 
The MDCTCodec employs powerful ConvNeXt v2 \cite{woo2023ConvNeXt} backbones for its encoder and decoder, and adopts RVQ for discretization, which performs multi-stage residual quantization to progressively refine feature representations and improve coding quality. 
The RVQ consists of $N$ $(N\ge 2)$ VQs, each with a codebook $\mathbb{W}_n = \{\bm{w}_{n,m} \in \mathbb{R}^{K} \mid m = 1, \dots, M\},n=1\dots,N$, where $K$ and $M$ denote codevector dimension and codebook size, respectively. 
When training MDCTCodec, we use \textit{single-speaker data} $\mathcal D_s=\{\bm{x} \}$ and retain the original adversarial loss, quantization loss, and spectral-level loss \cite{jiang2024mdctcodec}.



\vspace{-1mm}
\subsection{Base-Token Disentanglement Module}
\vspace{-1mm}

The BTD module consists of three submodules that take the mixed mel spectrogram as input to disentangle speakers’ base tokens. 

\begin{itemize}[leftmargin=*]
\item {\textbf{Mel Downsampling Block $\bm{\phi_{meld}}$:}} This submodule consists of $N_{meld}$ convolutional layers with stride $> 1$, aiming to further downsample the mel-spectrogram of the input mixed speech $\bm{y}$ along the temporal axis, so as to match the frame rate of the tokens quantized by MDCTCodec. 

\item {\textbf{Source-Intra Transformers $\bm{\phi_{intra}}$:}} Taking one-frame downsampled feature $\bm{z}_{meld}\in\mathbb{R}^{K_{meld}}$ as an example, it is processed by $N_{intra}$ source-intra Transformer blocks, yielding a same-dimensional feature $\bm{z}_{intra}\in\mathbb{R}^{K{meld}}$, i.e., $\bm{z}_{intra}=\phi_{intra}(\bm{z}_{meld})$. 
Each source-intra Transformer block is composed of \textit{self-attention} units, which operate on the mixed speech as a single source. 
Leveraging self-attention’s strength in modeling long-range dependencies, these blocks enhance the representation of the mixture. 

\item {\textbf{Anti-Consistency Source-Inter Transformers $\bm{\phi_{ac-inter}}$:}} This submodule consists of an anti-consistency bias generator (ACBG) $\phi_{ac}$ and $N_{inter}$ source-inter Transformers $\phi_{inter}$. 
The feature $\bm{z}_{intra}$ is first duplicated to form two sources, each added with trainable bias vectors $\bm{\delta}^{(1)},\bm{\delta}^{(2)}\in\mathbb R^{K_{meld}}$ generated by ACBG, which then serves as the input to $\phi_{inter}$. 
This bias acts as a perturbation for each source, enforcing anti-consistency by preventing inter-source uniformity and avoiding the model’s convergence to overly consistent solutions. 
Then, the two sources are respectively processed by source-inter Transformers composed of \textit{cross-attention} units to produce the corresponding results $\bm{z}_{inter}^{(1)},\bm{z}_{inter}^{(2)}\in\mathbb{R}^{K{meld}}$, i.e., 
\begin{equation}
\label{eq:z_inter}
\bm{z}_{inter}^{(1)},\bm{z}_{inter}^{(2)}=\phi_{inter}(\bm{z}_{intra}+\bm{\delta}^{(1)},\bm{z}_{intra}+\bm{\delta}^{(2)}).
\end{equation}
Leveraging the strength of cross-attention in modeling inter-source dependencies, these blocks enable more effective separation of the sources. 
Finally, $\bm{z}_{inter}^{(1)}$ and $\bm{z}_{inter}^{(2)}$ are respectively transformed by two linear layers followed by softmax, generating the probability distributions $\bm{p}_{base}^{(1)}, \bm{p}_{base}^{(2)} \in \mathbb{R}^M$, from which the disentangled base tokens are obtained via argmax sampling, i.e.,
\begin{equation}
\label{eq:base_token}
d_{base}^{(i)}=\arg\max_m(\bm{p}_{base}^{(i)}[1],\dots,\bm{p}_{base}^{(i)}[m],\dots,\bm{p}_{base}^{(i)}[M]),
\end{equation}
where $i=1,2$ and $d_{base}^{(i)}\in\{1,\dots,M \}$. 
\item {\textbf{Permutation-Invariant Cross-Entropy Training:}} Due to the unordered nature of base token disentanglement, we propose a permutation-invariant cross-entropy (PI-CE) loss to train the BTD module with \textit{mixture data} $\mathcal D_m=\{(\bm{y},\bm{x}_1,\bm{x}_2) \}$. 
Suppose the target speeches $\bm{x}_1$ and $\bm{x}_2$ produce quantized tokens $d_1^{(1)},d_1^{(2)}\in\{1,\dots,M \}$ from the first VQ of MDCTCodec, respectively. 
The PI-CE loss is defined between the one-hot probability distributions of the quantized tokens and the probability distributions of the disentangled base tokens, specifically as:
\begin{equation}
\label{eq:3}
    \mathcal{L}_{PI-CE}=-\mathbb E_{\mathcal D_m}\min_{\pi\in \mathbb S_2} \sum_{i=1}^2 \sum_{j=1}^2 \bm{1}_{\{j=\pi[i]\}}\ln(\bm{p}_{base}^{(i)}[d_1^{(j)}]),
\end{equation}
where $\mathbb S_2=\{(1, 2),(2,1) \}$ is a two-element set composed of ordered pairs, and $\bm{1}_{\{condition\}}$ denotes the indicator function, which equals 1 if the condition is satisfied and 0 otherwise.

\end{itemize}

\vspace{-3mm}
\subsection{Auxiliary-Token Serial Prediction Module}
\vspace{-1mm}

The ATSP modules, based on the base tokens disentangled by the BTD module, predict the remaining auxiliary tokens for each source in a parallel and independent yet serial manner. 
\begin{itemize}[leftmargin=*]
\item {\textbf{Token Serial Prediction Process:}} Taking the $i$-th ATSP branch as an example, it consists of $N-1$ sub-predictors that perform sequential token probability prediction, where each prediction is conditioned on the previously predicted results. 
Suppose the auxiliary token sequence is $\bm{d}^{(i)}_{aux}=[d^{(i)}_{aux,1},\dots,d^{(i)}_{aux,N-1}]$, then the objective of this branch is to model the following probability distribution: 
\begin{equation}
\label{eq:4}
p(\bm{d}^{(i)}_{aux})=\prod_{n=1}^{N-1}p(d^{(i)}_{aux,n}|d^{(i)}_{aux,1},\dots,d^{(i)}_{aux,n-1},d^{(i)}_{base}).
\end{equation}
Specifically, each sub-predictor consists of $N_{lstm}$ long short-term memory (LSTM) layers and $N_{con}$ Conformer blocks, with the input being the sum of the previously predicted results, i.e.,
\begin{equation}
\label{eq:5}
\scalebox{1}{$
\bm{z}^{(i)}_{aux,n}=\mathbb L(d^{(i)}_{base},\mathbb W_1)+ \bm{1}_{\{n\ge 2\}}\sum\limits_{n'=1}^{n-1} \mathbb L(d^{(i)}_{aux,n'},\mathbb W_{n'+1}),
$}
\end{equation}
where $n = 1, \dots, N-1$ denotes the index of the sub-predictors, and $\mathbb{L}$ denotes codebook lookup operation. 
Similar to base token prediction, the output of each sub-predictor is passed through a softmax layer to generate a probability distribution $\bm{p}_{aux,n}^{(i)}\in \mathbb{R}^M$, from which the auxiliary token $d_{aux,n}^{(i)}$ is derived via argmax sampling. 
The above process iterates sequentially from $n=1$ to $N-1$ until the complete auxiliary token sequence $\bm{d}^{(i)}_{aux}$ is generated. 

\item {\textbf{Teacher-Forcing-based Cross-Entropy Training:}} We adopt a teacher-forcing (TF) strategy to train an ATSP branch with \textit{single-speaker data} $\mathcal D_s=\{\bm{x} \}$ to reduce training difficulty, while the parallel branches share weights. 
In this strategy, each sub-predictor takes the previous quantization results of MDCTCodec as input (see the red dashed arrows in Fig. \ref{fig:2}) to generate a probability distribution $\tilde{\bm{p}}_{aux,n}\in \mathbb{R}^M$. 
The teacher-forcing-based cross-entropy loss for training is defined as
\begin{equation}
\label{eq:6}
\mathcal{L}_{TF-CE}=-\mathbb E_{\mathcal D_s} \sum_{n=1}^{N-1} \ln(\tilde{\bm{p}}_{aux,n}[d_{n+1}]),
\end{equation}
where $d_{n+1}\in\{1,\dots,M \}$ is the token quantized from $\bm{x}$ by the $(n+1)$-th VQ in MDCTCodec. 

\end{itemize}

Finally, the base token $d_{base}^{(i)}$ from the BTD module and the auxiliary tokens $\bm{d}^{(i)}_{aux}$ from the ATSP module are codebook-looked up and added, and the resulting embeddings are fed into MDCTCodec’s decoder to reconstruct the separated speeches $\hat{\bm{x}}_i$, $i=1,2$.




\vspace{-2mm}
\section{Experiments and Results}
\vspace{-1mm}
\label{sec:prior}
\subsection{Experimental Setup}
\vspace{-1mm}
Libri2Mix-clean \cite{cosentino2020librimix} is used as the mixture data $\mathcal D_m$ in our experiments, where all speech utterances are sampled at 16 kHz. 
It consists of a training set, a development set, and a test set, respectively containing 270, 11, and 11 hours of 2-speaker mixed utterances along with the corresponding single-speaker utterances. 
These single-speaker utterances are merged into single-speaker data $\mathcal D_s$ for experiments.


For CodeSep\footnote{Speech samples can be found at: https://redmist328.github.io/CodeSep/.}, its MDCTCodec configuration fully follows \cite{jiang2024mdctcodec}, using $N=4$ VQs with a codebook size of $M=1024$ and a codevector dimension of $K=32$. 
In the BTD module, $\phi_{meld}$ extracts an 80-dimensional mixed mel-spectrogram with a frame shift of 80, which is then passed through $N_{meld}=3$ convolutional layers with stride $=2$ to produce features of dimension $K_{meld}=256$. 
$\phi_{intra}$ and $\phi_{ac\text{-}inter}$ have the same number of Transformers, i.e., $N_{Intra}=N_{Inter}=4$. 
In the ATSP modules, each sub-predictor adopts $N_{lstm}=2$ LSTM layers and $N_{con}=3$ Conformer blocks. 
All the modules were trained using the AdamW optimizer \cite{kingma2014adam} for up to 1 million steps. 
Attributable to the design of CodeSep, transmission or storage of only the base tokens suffices to reconstruct the separated speech. 
Under this setting, the bitrate of CodeSep is merely 1 kbps. 

The two separation–compression solutions at multiple bitrates described in Section \ref{sec: task def} serve as baselines for comparison with CodeSep at 1 kbps. 
Assume the preset bitrate of the baselines is $b$ kbps. 
For the FCTS solution, the mixed speech is first compressed with an $b$ kbps MDCTCodec and then separated by Sepformer \cite{subakan2021attention}; whereas for the FSTC solution, the mixed speech is first separated by Sepformer \cite{subakan2021attention} and each resulting speech is then compressed with an $b/2$ kbps MDCTCodec. 
We also include the separation-only Sepformer \cite{subakan2021attention} (interpreted as infinite bitrate) purely as an upper-bound reference rather than for direct comparison.

\vspace{-3mm}
\subsection{Evaluation Metrics}
\vspace{-1mm}

We employed both objective and subjective metrics to comprehensively evaluate the performance of CodeSep. 
For objective evaluation, we use two non-intrusive metrics, i.e., UTMOS \cite{saeki2022utmos} and DNSMOS \cite{reddy2021dnsmos}, to assess the quality of the separated speech. 
As this work does not target pure speech separation, we do not employ the commonly used SI-SNRi and SI-SDRi metrics \cite{subakan2021attention} in that domain. 
Moreover, as suggested by \cite{yip2024towards,yip2024speech}, intrusive metrics, e.g., SI-SNRi and SI-SDRi, are unsuitable for codec-based speech separation, likely because token-level modeling is indirect compared with direct speech prediction.


To address the limitations of objective evaluation, we carried out comprehensive subjective experiments. 
All subjective experiments were conducted on the Amazon Mechanical Turk (AMT) platform, with feedback collected from no fewer than 40 native English speakers. 
First, we conducted naturalness mean opinion score (NMOS) and similarity mean opinion score (NMOS) tests to compare CodeSep with the baselines at the same bitrate (i.e., 1 kbps).
Each MOS test batch included 20 generated test set's speech samples per solution, accompanied by natural speech samples as references. 
Listeners rated each sample on a 5-point scale with 0.5-point increments for naturalness and speaker similarity, corresponding to NMOS and SMOS, respectively. 
Then, we carried out naturalness ABX (NABX) and similarity ABX (SABX) tests, comparing CodeSep at 1 kbps with the baselines at higher bitrates in a pairwise manner. 
Each ABX test batch likewise included 20 generated speech samples per solution, with natural speech samples as references.
Listeners were asked to judge, within each pair, which speech sample had better naturalness (for NABX) or better speaker similarity (for SABX), or to indicate no preference (N/P) between the two. 
We reported average preference scores and $t$-test $p$-values for statistical significance.

\begin{table}[t!]
	\centering
	\caption{Objective metrics and subjective MOS test results with 95\% confidence intervals of the proposed CodeSep and baselines at 1 kbps on the LibriMix test set.
}\label{tab1}
\setlength{\arrayrulewidth}{1pt}
	\adjustbox{width=0.48\textwidth}{
		\renewcommand{\arraystretch}{0.97}\setlength{\tabcolsep}{0.5mm}{
		\begin{tabular}{c c|c c|c c}
			\hline

            \hline
        &Bitrate &UTMOS$\uparrow$&DNSMOS$\uparrow$&NMOS$\uparrow$&SMOS$\uparrow$\\
			\hline
            \textbf{CodeSep} &1 kbps&\textbf{3.14}&\textbf{3.67}&\textbf{3.65 ($\pm$0.08)}&\textbf{3.43 ($\pm$ 0.09)}\\
			\hline
			\textbf{FCTS}&1 kbps&1.34 &3.03&2.96 ($\pm$0.09)&2.86 ($\pm$0.09)\\

			\textbf{FSTC} &1 kbps&1.99&3.33&3.24 ($\pm$0.09)&3.15 ($\pm$0.09)\\
            \hline
			\textbf{Sepformer}&$+\infty$&3.54&3.55&-&-\\

            \hline
            
            \hline
	\end{tabular}}}
\end{table}

\addtolength{\textfloatsep}{-0.7cm}

\vspace{-6mm}
\subsection{Comparison at the Low-Bitrate Condition}
\vspace{-1mm}
We first compared the proposed CodeSep with the baseline FCTS and FSTC solutions at the same low bitrate of 1 kbps (i.e., $b=1$) for fairness. 
The objective and subjective results are presented in Table \ref{tab1}, where Sepformer is included only as an upper-bound reference with objective metrics reported. 
We can see whether in objective metrics reflecting speech quality (i.e., UTMOS and DNSMOS) or in subjective NMOS and SMOS reflecting speech naturalness and speaker similarity, our CodeSep significantly ($p<0.01$) outperformed both the FCTS and FSTC solutions, which merely combine speech separation and speech codec modules in a simple manner. 
The FCTS solution directly compresses the mixed speech, a challenging task that imposes high demands on the codec’s bitrate and modeling capacity. 
Consequently, the mixed speech reconstructed after low-bitrate compression suffers from poor quality, which in turn degrades the separated speech. 
Although the FSTC solution first performs speech separation and thus yields high-quality separated speech, under low-bitrate conditions the bitrate for each separated stream is halved, leading to a sharp drop in reconstruction quality. 
In contrast, our CodeSep adopts a JSAC solution, where the BTD module jointly separates and compresses mixed speech at the token level, while the ATSP module further enhances separation quality without adding to the bitrate, thereby achieving high-quality speech separation under low bitrate conditions. 
In addition, with Sepformer (equivalent to infinite bitrate) as the upper-bound reference, our CodeSep even shows an advantage in the DNSMOS metric, further demonstrating its strong effectiveness.

\vspace{-5mm}
\subsection{Comparison against Higher-Bitrate Baselines}
\vspace{-3mm}
\begin{table}[t!]
	\centering
	\caption{Objective metrics of the proposed CodeSep at 1 kbps and FSTC solution at higher bitrates on the LibriMix test set.
}\label{tab2}
\setlength{\arrayrulewidth}{0.7pt}
		\renewcommand{\arraystretch}{0.9}
		\begin{tabular}{c | c| c c}
    \hline
    
    \hline
        &Bitrate  &UTMOS$\uparrow$&DNSMOS$\uparrow$\\
			\hline
            \textbf{CodeSep} &1 kbps&\textbf{3.14}&\textbf{3.67}\\
            \hline
			\textbf{FSTC}&2 kbps&2.30& 3.44\\
			\textbf{FSTC}&4 kbps&2.87&3.53\\
			\textbf{FSTC} &8 kbps&3.11&3.56\\
            \hline

			\hline
	\end{tabular}
\end{table}

\begin{table}[t!]
\centering
    \caption{Subjective ABX preference scores of the proposed CodeSep at 1 kbps and FSTC solution at higher bitrates.}  
    \setlength{\arrayrulewidth}{1pt}
    \label{tab3}  
	\adjustbox{width=0.48\textwidth}{
		\renewcommand{\arraystretch}{1.1}\setlength{\tabcolsep}{1.9mm}{
    \begin{tabular}{c|c|c c c c c}
        \hline  

        \hline
        & Bitrate & \textbf{CodeSep}&\textbf{FSTC} &N/P&$p$-value \\
        \hline
                \multirow{3}{*}{\makecell{NABX}} 
        & 1 vs. 2 kbps& \textbf{55.83\%}&41.90\%&2.27\%&$<$0.01 \\  
        & 1 vs. 4 kbps &\textbf{ 52.77\%}&42.97\%&4.26\%&$<$0.01 \\  
        & 1 vs. 8 kbps & 38.57\%&\textbf{53.57\%}&7.86\%&$<$0.01 \\  
        \hline
        \multirow{3}{*}{\makecell{SABX} }
        & 1 vs. 2 kbps & \textbf{54.29\%}&41.79\%&3.94\% &$<$0.01\\  
        & 1 vs. 4 kbps & 47.23\%&46.91\%&5.86\%&0.78 \\  
        & 1 vs. 8 kbps &45.43\%&44.00\%&10.57\%&0.68 \\  
        \hline

        \hline
    \end{tabular}}}
\end{table}
\begin{table}[t!]
\centering
    \caption{Subjective ABX preference scores of the proposed CodeSep at 1 kbps and its ablated variants.}  
    \setlength{\arrayrulewidth}{1pt}
    \label{tab:4}  
	\adjustbox{width=0.48\textwidth}{
        {
    \begin{tabular}{c|c c c c c}
        \hline  

        \hline
         & \textbf{CodeSep}&\textbf{w/o ACBG}&\textbf{w/o TF} &N/P&$p$-value \\
                 \hline
        \multirow{2}{*}{\makecell{NABX}} 
        & {48.72\%}&42.44\%&-&8.85\%&0.066 \\  
       &\textbf{55.19\%}&-&27.22\%& 17.59\%&$<$0.01 \\  
        \hline
        \multirow{2}{*}{\makecell{SABX} }
        & \textbf{54.23\%}&38.33\%&-&7.44\% &$<$0.01\\  
        & 43.15\%&-&37.41\%&19.44\%&0.14 \\  

        \hline

        \hline
    \end{tabular}}}
\end{table}
    
    
    
    

To assess the bitrate savings of the JSAC solution proposed in CodeSep, we compared CodeSep at 1 kbps with baselines at higher bitrates. 
Only FSTC at $b=2,4,8$ kbps is considered as the baseline, since it clearly outperformed FCTS at the same bitrate in Table \ref{tab1}. 
Interestingly, the results in Table \ref{tab2} reveal that CodeSep, operating at only 1 kbps, attained higher UTMOS and DNSMOS scores than the FSTC solution across all three higher bitrate settings. 
As shown in Table \ref{tab3}, CodeSep at 1 kbps yielded significantly better naturalness in the NABX test than FSTC at 2 and 4 kbps, though the advantage reversed at 8 kbps. 
This suggests that CodeSep saved at least 3 kbps while maintaining naturalness of separated speech, thus improving transmission or storage efficiency. 
In the SABX test, FSTC should theoretically yield higher speaker similarity, but its 2 kbps performance is unsatisfactory, likely due to degraded timbre from low-bitrate coding. 
As the bitrate increases, FSTC’s speaker similarity improved markedly, and our CodeSep at 1 kbps even matched FSTC at 8 kbps, confirming the strength of our separation capability.

\vspace{-3mm}
\subsection{Ablation Studies}
\vspace{-1mm}
Finally, we ablated the ACBG in the BTD module (i.e., w/o ACBG) and the teacher-forcing strategy in the ATSP module (i.e., w/o TF) to verify their effectiveness. 
The subjective NABX and SABX results are shown in Table \ref{tab:4}. 
Interestingly, removing ACBG did not affect speech naturalness but clearly reduced speaker similarity. 
This suggests that the random perturbations from ACBG enlarged inter-source distribution differences, thereby facilitating separation. 
In contrast, ablating the teacher-forcing strategy yielded the opposite effect, indicating that the serial training structure of the ATSP module without teacher forcing increases learning difficulty and degrades the naturalness of the separated speech. 
Since the two ablated components belong to different modules, the above experiments also reveal their respective roles, with the BTD module responsible for separation and the ATSP module improving speech quality.

\vspace{-3mm}
\section{Conclusion}
\vspace{-3mm}
This paper investigated the integrated task of speech separation and compression. 
For this task, we introduced CodeSep, a novel codec-driven speech separation model that combines a plain codec, a BTD module, and parallel ATSP modules. 
The CodeSep disentangles mixed speech mel-spectrogram into base tokens for efficient transmission or storage and further refines them with auxiliary tokens to improve separation quality, achieving joint speech separation and low-bitrate compression. 
Both objective and subjective show that CodeSep is able to preserve high speaker similarity and naturalness in separated speech even at a low bitrate of just 1 kbps. 
In future work, we will explore separation–compression scenarios with more mixed speakers and further optimize the architecture of CodeSep.


\bibliographystyle{IEEEbib}
\bibliography{strings,refs}

@article{luo2019conv,
  title={Conv-tasnet: Surpassing ideal time--frequency magnitude masking for speech separation},
  author={Luo, Yi and Mesgarani, Nima},
  journal={IEEE/ACM transactions on audio, speech, and language processing},
  volume={27},
  number={8},
  pages={1256--1266},
  year={2019},
}

@inproceedings{jiang2024mdctcodec,
  title={Mdctcodec: A lightweight mdct-based neural audio codec towards high sampling rate and low bitrate scenarios},
  author={Jiang, Xiao-Hang and Ai, Yang and Zheng, Rui-Chen and Du, Hui-Peng and Lu, Ye-Xin and Ling, Zhen-Hua},
  booktitle={Porc. SLT},
  pages={540--547},
  year={2024},
}

@inproceedings{chen2020dual,
  title={Dual-Path Transformer Network: Direct Context-Aware Modeling for End-to-End Monaural Speech Separation},
  author={Chen, Jingjing and Mao, Qirong and Liu, Dong},
  booktitle={Proc. Interspeech},
  pages={2642--2646},
  year={2020}
}

@inproceedings{chen2023neural,
  title={A Neural State-Space Modeling Approach to Efficient Speech Separation},
  author={Chen, Chen and Yang, Chao-Han Huck and Li, Kai and Hu, Yuchen and Ku, Pin-Jui and Chng, Eng Siong},
  booktitle={Proc. Interspeech},
  pages={3784--3788},
  year={2023}
}

@inproceedings{reddy2021dnsmos,
  title={{DNSMOS}: A non-intrusive perceptual objective speech quality metric to evaluate noise suppressors},
  author={Reddy, Chandan KA and Gopal, Vishak and Cutler, Ross},
  booktitle={Proc. ICASSP},
  pages={6493--6497},
  year={2021},
}

@article{zeghidour2021soundstream,
  title={Soundstream: An end-to-end neural audio codec},
  author={Zeghidour, Neil and Luebs, Alejandro and Omran, Ahmed and Skoglund, Jan and Tagliasacchi, Marco},
  journal={IEEE/ACM Transactions on Audio, Speech, and Language Processing},
  volume={30},
  pages={495--507},
  year={2021},
}

@inproceedings{saeki2022utmos,
	title={{UTMOS: Utokyo-sarulab system for voiceMOS Challenge 2022}},
	author={Saeki, Takaaki and Xin, Detai and Nakata, Wataru and Koriyama, Tomoki and Takamichi, Shinnosuke and Saruwatari, Hiroshi},
	booktitle={Proc. Interspeech},
	year={2022},
  pages     = {4521--4525},
}

@inproceedings{kingma2014adam,
	title={Adam: A method for stochastic optimization},
	author={Kingma, Diederik P and Ba, Jimmy},
	booktitle={Proc. ICLR},
	year={2015}
}

@inproceedings{woo2023ConvNeXt,
	title={{ConvNeXt v2: Co-designing and scaling convnets with masked autoencoders}},
	author={Woo, Sanghyun and Debnath, Shoubhik and Hu, Ronghang and Chen, Xinlei and Liu, Zhuang and Kweon, In So and Xie, Saining},
	booktitle={Proc. CVPR},
        pages={16133--16142},
	year={2023}
}

@article{kumar2023high,
  title={High-fidelity audio compression with improved rvqgan},
  author={Kumar, Rithesh and Seetharaman, Prem and Luebs, Alejandro and Kumar, Ishaan and Kumar, Kundan},
  journal={Advances in Neural Information Processing Systems},
  volume={36},
  pages={27980--27993},
  year={2023}
}

@inproceedings{yang2022tfpsnet,
  title={TFPSNet: Time-frequency domain path scanning network for speech separation},
  author={Yang, Lei and Liu, Wei and Wang, Weiqin},
  booktitle={Proc. ICASSP},
  pages={6842--6846},
  year={2022},
}

@inproceedings{subakan2021attention,
  title={Attention is all you need in speech separation},
  author={Subakan, Cem and Ravanelli, Mirco and Cornell, Samuele and Bronzi, Mirko and Zhong, Jianyuan},
  booktitle={Proc. ICASSP},
  pages={21--25},
  year={2021},
}

@inproceedings{yip2024speech,
  title={Speech separation using neural audio codecs with embedding loss},
  author={Yip, Jia Qi and Kwok, Chin Yuen and Ma, Bin and Chng, Eng Siong},
  booktitle={Proc. APSIPA ASC},
  pages={1--6},
  year={2024},
}

@inproceedings{li2023design,
  title={On the design and training strategies for rnn-based online neural speech separation systems},
  author={Li, Kai and Luo, Yi},
  booktitle={Proc. ICASSP},
  pages={1--5},
  year={2023},
}

@inproceedings{luo2020dual,
  title={Dual-path rnn: efficient long sequence modeling for time-domain single-channel speech separation},
  author={Luo, Yi and Chen, Zhuo and Yoshioka, Takuya},
  booktitle={Proc. ICASSP},
  pages={46--50},
  year={2020},
}

@inproceedings{li2022use,
  title={On the Use of Deep Mask Estimation Module for Neural Source Separation Systems},
  author={Li, Kai and Hu, Xiaolin and Luo, Yi},
  booktitle={Proc. Interspeech},
  pages={5328--5332},
  year={2022}
}

@article{hu2021speech,
  title={Speech separation using an asynchronous fully recurrent convolutional neural network},
  author={Hu, Xiaolin and Li, Kai and Zhang, Weiyi and Luo, Yi and Lemercier, Jean-Marie and Gerkmann, Timo},
  journal={Advances in Neural Information Processing Systems},
  volume={34},
  pages={22509--22522},
  year={2021}
}

@article{li2022efficient,
  title={An efficient encoder-decoder architecture with top-down attention for speech separation},
  author={Li, Kai and Yang, Runxuan and Hu, Xiaolin},
  journal={arXiv preprint arXiv:2209.15200},
  year={2022}
}

@inproceedings{tzinis2020sudo,
  title={Sudo rm-rf: Efficient networks for universal audio source separation},
  author={Tzinis, Efthymios and Wang, Zhepei and Smaragdis, Paris},
  booktitle={Proc. MLSP},
  pages={1--6},
  year={2020},
}

@inproceedings{yip2024towards,
  title={Towards Audio Codec-based Speech Separation},
  author={Yip, Jia Qi and Zhao, Shengkui and Ng, Dianwen and Chng, Eng Siong and Ma, Bin},
  booktitle={Proc. Interspeech},
  pages={2190--2194},
  year={2024}
}

@article{cosentino2020librimix,
  title={Librimix: An open-source dataset for generalizable speech separation},
  author={Cosentino, Joris and Pariente, Manuel and Cornell, Samuele and Deleforge, Antoine and Vincent, Emmanuel},
  journal={arXiv preprint arXiv:2005.11262},
  year={2020}
}

@article{li2024spmamba,
  title={Spmamba: State-space model is all you need in speech separation},
  author={Li, Kai and Chen, Guo and Yang, Runxuan and Hu, Xiaolin},
  journal={arXiv preprint arXiv:2404.02063},
  year={2024}
}

@inproceedings{le2019sdr,
  title={SDR--half-baked or well done?},
  author={Le Roux, Jonathan and Wisdom, Scott and Erdogan, Hakan and Hershey, John R},
  booktitle={Proc. ICASSP},
  pages={626--630},
  year={2019},
}

@article{cherry1953some,
  title={Some experiments on the recognition of speech, with one and with two ears},
  author={Cherry, Edward Collin},
  journal={Journal of the acoustical society of America},
  volume={25},
  pages={975--979},
  year={1953}
}

\end{document}